\documentclass[%
 aip,
 jmp,%
 amsmath,amssymb,
 reprint,%
]{revtex4-1}

\usepackage{graphicx}
\usepackage{dcolumn}
\usepackage{bm}
\usepackage{color}
\usepackage{multirow} 

\begin{document}

\preprint{AIP/123-QED}

\title{Temperature dependent transport characteristics of graphene/n-Si diodes}

\author{S. Parui}
\thanks{Present address: CIC nanoGUNE Consolider, Tolosa Hiribidea 76, 20018 Donostia-San Sebastian, Basque Country, Spain}
\affiliation{ 
Physics of Nanodevices, Zernike Institute for Advanced Materials, University of Groningen, Nijenborgh 4, 9747 AG Groningen, The Netherlands
}%

\author{R. Ruiter}
\author{P. J. Zomer}
\author{M. Wojtaszek}
\author{B. J. van Wees}
\author{T. Banerjee}
\email{T.Banerjee@rug.nl}
\affiliation{ 
Physics of Nanodevices, Zernike Institute for Advanced Materials, University of Groningen, Nijenborgh 4, 9747 AG Groningen, The Netherlands
}%
\date{\today}

\begin{abstract}
Realizing an optimal Schottky interface of graphene on Si is challenging, as the electrical transport strongly depends on the graphene quality and the fabrication processes. Such interfaces are of increasing research interest for integration in diverse electronic devices as they are thermally and chemically stable in all environments, unlike standard metal/semiconductor interfaces. We fabricate such interfaces with n-type Si at ambient conditions and find their electrical characteristics to be highly rectifying, with minimal reverse leakage current ($<$10$^{-10}$ A) and rectification of more than $10^6$. We extract Schottky barrier height of 0.69 eV for the exfoliated graphene and 0.83 eV for the CVD graphene devices at room temperature. The temperature dependent electrical characteristics suggest the influence of inhomogeneities at the graphene/n-Si interface. A quantitative analysis of the inhomogeneity in Schottky barrier heights is presented using the potential fluctuation model proposed by Werner and G\"{u}ttler.
 
\end{abstract}

\maketitle
\section{\label{sec:level1}Introduction}  
Schottky interfaces between graphene (Gr) and semiconductors are relevant for future applications in electronic devices such as solar cells, high speed logic gates, photodetectors, three-terminal transistors, hot electron based devices etc. \cite{Barristor, Solarcell, Chen, TongayPRX, Photodectors, Subir_PhD}. A key requisite is to fabricate such Schottky interfaces with low leakage current in reverse bias and a high current in forward bias in order to obtain stable and highly rectifying diodes \cite{Sze, Rhoderick}. Recently, a three-terminal ``graphene-barristor" was demonstrated based on tunable Schottky barrier across electrostatically gated Gr/Si interface \cite{Barristor}, using large area chemical vapor deposition (CVD) grown graphene. Although CVD graphene interfaces are found to have an advantage over their exfoliated counterparts due to their large area and easier fabrication methods, Gr/Si Schottky interfaces using exfoliated graphene have also been fabricated \cite{Chen}. Usually for both types of interfaces made by either exfoliated or CVD graphene, the rectification, which is the ratio between forward saturation current and reverse saturation current ($I_{FS}/I_{RS}$) is reported to be quite low ($\sim$10$^3$) and accompanied by a large ideality factor, $\eta$ of $\sim$5 to $\sim$33, primarily in exfoliated graphene diodes at room temperature \cite{Chen}. Such a low value of the rectification can arise due to the weak bonding between graphene and Si during fabrication causing high reverse leakage current and reducing the diode quality.
   
\begin{figure}
\includegraphics[scale=0.69]{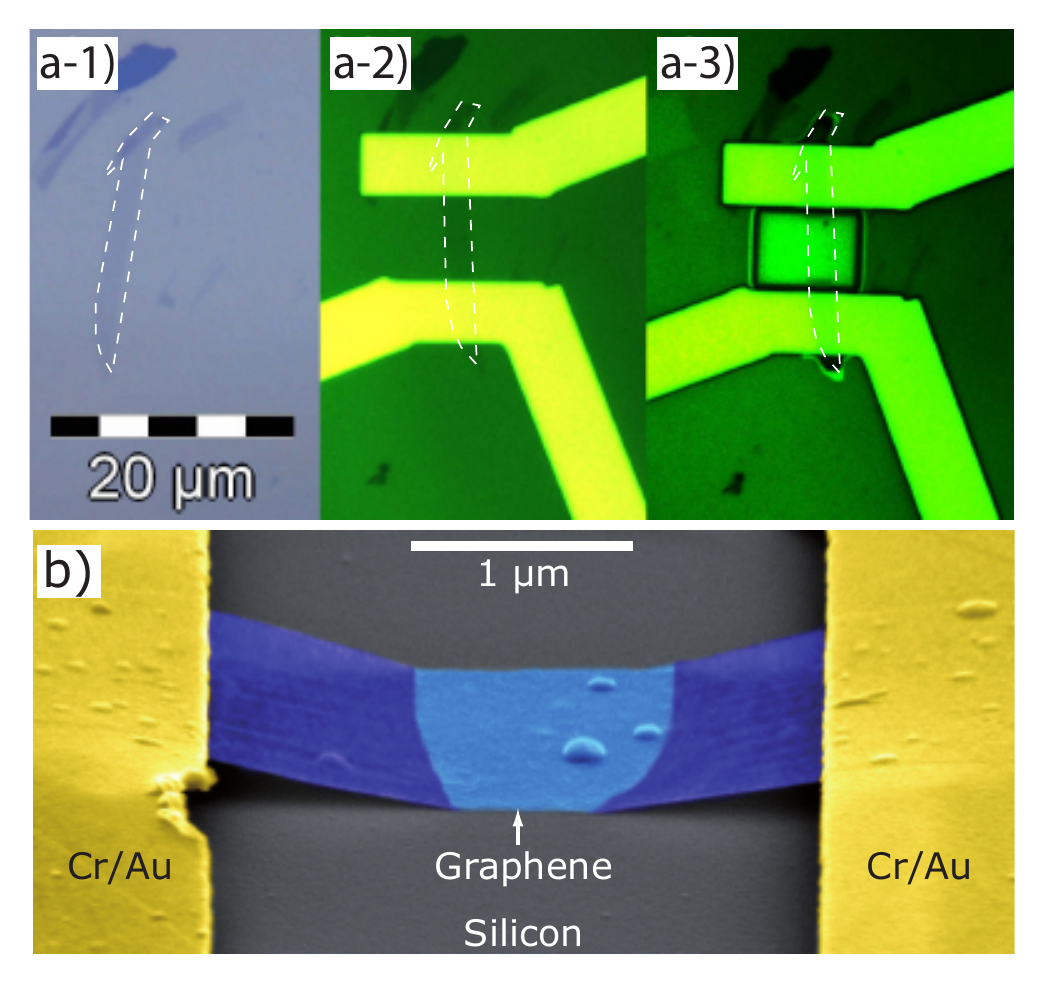}
\caption{\label{fig:Device} (a-1) Graphene flake on the SiO$_2$ surface. (a-2) The flake is connected by Cr/Au contacts. (a-3) The device after etching of the SiO$_2$ underneath of the graphene. (b) Scanning electron micrograph (SEM) of a typical exfoliated-Gr/Si device (false color). 
}
\end{figure}
 
\begin{figure}[tbp]
\includegraphics[scale=0.69]{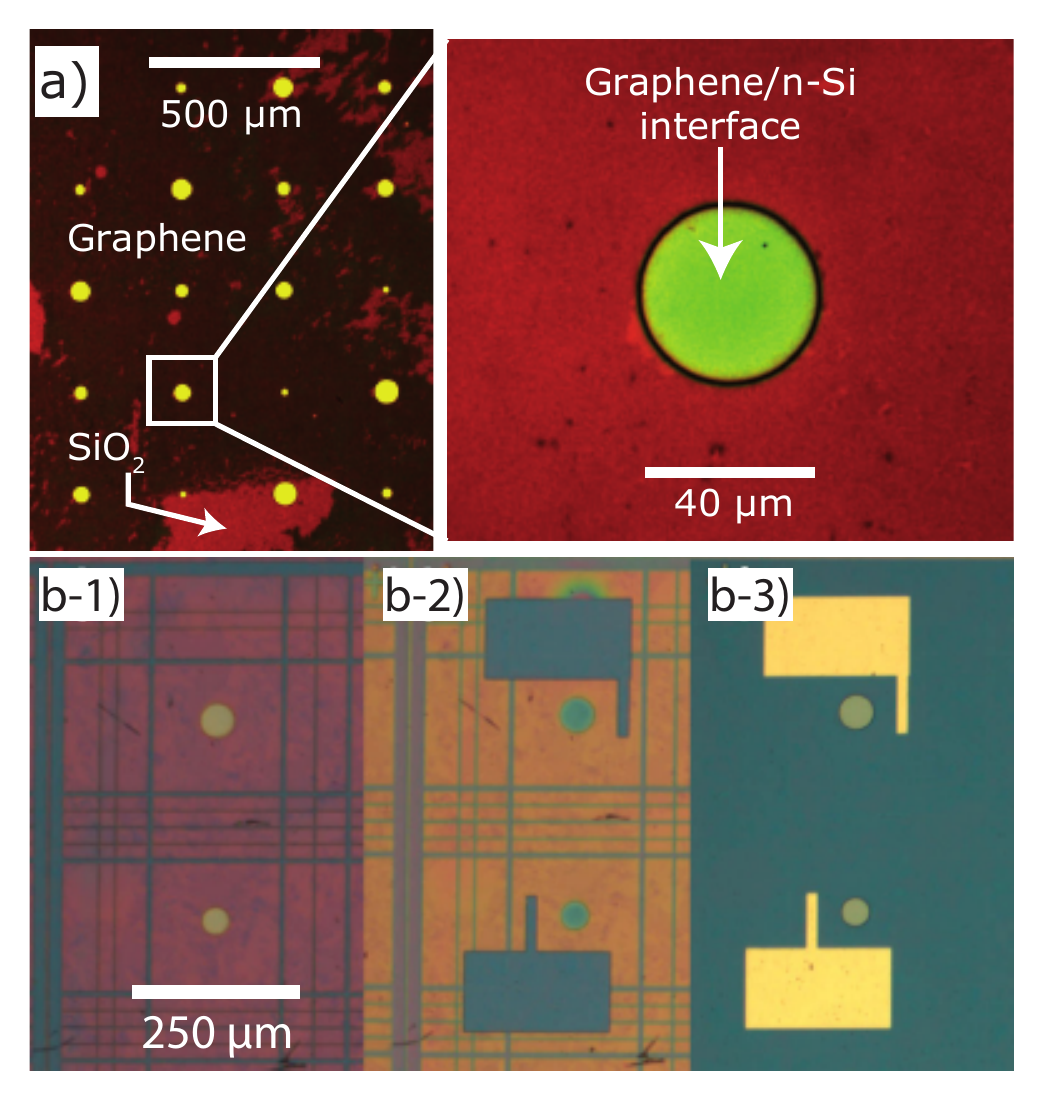}
\caption{\label{fig:IV} (a) Optical image of the transferred CVD graphene on the patterned substrate and the zoomed view of one such device. (b-1) Different device areas are defined by e-beam lithography and etched in oxygen plasma. (b-2) Another layer of resist is added on top where the contact areas are defined and (b-3) then Ti/Au contacts are evaporated. 
}
\end{figure}

In this work, we demonstrate highly-rectifying Gr/Si interfaces using both exfoliated and large area CVD graphene on n-type Si with a doping concentration $N_d = 10^{15}$ cm$^{-3}$ and compare their electrical transport. We analyse the temperature dependence of the diode characteristics and find that features in the forward bias regime can be associated with inhomogeneities at the Gr/n-Si interfaces arising due to potential fluctuations in graphene because of ripples \cite{Rajput}, relatively high conduction at the edge and graphene grain boundaries \cite{Tsen}. An understanding of the temperature dependence of transport for both types of Gr/n-Si interface will be important for their implementation in diverse electronic devices.

\section{\label{sec:level1}Results and Discussion}
For the exfoliated-Gr/Si device, the graphene flake is exfoliated on top of a n-Si/SiO$_2$ (300 nm) substrate as shown in Fig.~\ref{fig:Device}(a-1) and is clamped by two Cr(6 nm)/Au(35 nm) contacts on the sides of the flake [Fig.~\ref{fig:Device}(a-2)] using e-beam lithography followed by deposition using e-beam evaporation. A second lithography step opens up a window in the PMMA mask in between the two electrodes. The substrate with the clamped graphene layer is then dipped in buffered hydrofluoric acid (BHF) to remove the 300 nm of SiO$_2$, followed by immersing it in deionized (DI) water. This leads to a contact between the Si surface and the clamped graphene when the substrate is taken out of the DI water and left to dry. We noticed that the graphene flake becomes non-suspended and forms a robust contact with Si when the electrode separation is more than 2 $\mu$m and can withstand further cleaning of the PMMA by hot acetone and isopropanol as shown in Fig.~\ref{fig:Device}(a-3) and Fig.~\ref{fig:Device}(b). Fig.~\ref{fig:Device}(b) represents the scanning electron microscopy image of a typical Gr/Si interface taken at 70$^\circ$ angle with respect to the surface of the substrate.

The CVD-Gr/Si diodes are fabricated by a different approach using a polydimethylsiloxane (PDMS) stamp to transfer graphene from Cu foil \cite{Lee} onto a patterned substrate. The Cu foil is etched away in FeCl$_3$ aqueous solution, leaving the graphene on a PDMS-stamp support. Using e-beam lithography, we pattern circular areas in PMMA mask on a Si/SiO$_2$ substrate and etch SiO$_2$ in BHF. After removing the PMMA mask, the substrate is submerged in 1\% hydrofluoric acid (HF) for H-termination followed by the transfer of CVD graphene using the PDMS stamp (shown in Fig.~\ref{fig:IV}(a)). Subsequently, the graphene in between the different diode areas is disconnected by oxygen plasma etching [see Fig.~\ref{fig:IV}(b-1)] and then Ti (6 nm) and Au (35 nm) are deposited [see Fig.~\ref{fig:IV}(b-2) and (b-3)] to contact graphene. 

\begin{figure}
\includegraphics[scale=0.79]{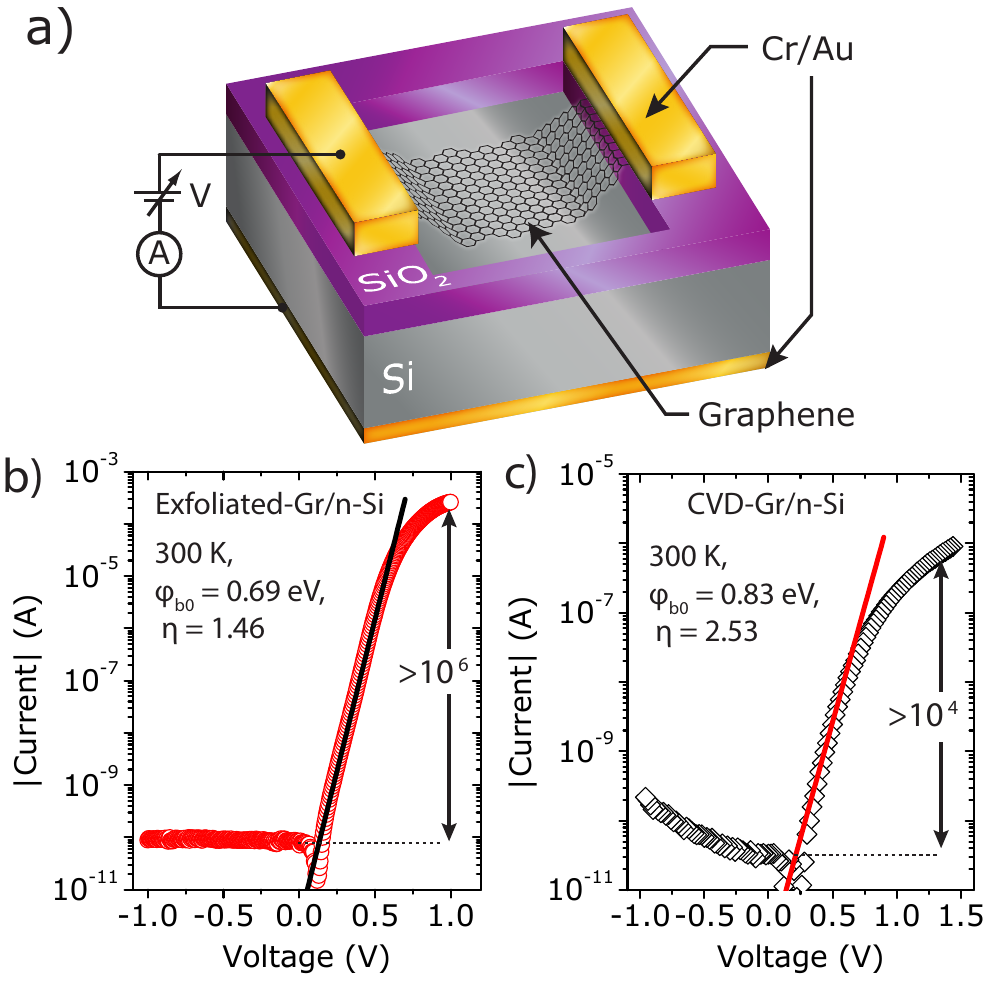}
\caption{\label{fig:RT_I-V}(a) The fabricated device schematic for a two-terminal electrical {\it I-V} measurement. (b, c) The {\it I-V} characteristics of the exfoliated-graphene device (b) and the CVD-graphene device (c) are measured at room temperature. The straight lines are fitted using the thermionic emission equation. 
}
\end{figure}

Figure~\ref{fig:RT_I-V}(a) represents the two-terminal electrical {\it I-V} measurement schematic. The measurements are performed in dark using a variable temperature cryostat. The room temperature (RT) diode characteristics of the exfoliated Gr/Si interface show low reverse leakage current ($<$10$^{-10}$ A) [see Fig.~\ref{fig:RT_I-V}(b)]. The forward current is measured to be $>$10$^{-4}$ A at 1 V for an optimized diode, with a large linear region as can be seen in the $\log(I)${\it-V} plot, and a rectification of $I_{FS}/I_{RS}>$ 10$^6$ at RT. From the {\it I-V} characteristics of CVD Gr/Si interface, the forward saturated current is measured to be $>$10$^{-6}$ A at 1.5 V [see Fig.~\ref{fig:RT_I-V}(c)] whereas the reverse saturation current is similar to that of the exfoliated device. Although the interface area between CVD-graphene and Si is much larger ($\sim$1250 $\mu$m$^2$) than between the exfoliated graphene and Si ($\sim$10.7 $\mu$m$^2$) as determined from the atomic force microscopy image of respective devices, the forward saturated current is lower. This can be explained by the lower transport quality of the CVD-graphene itself and by the Gr/Si interface which is cleaner for the exfoliated graphene. We must note that the CVD-graphene devices were annealed overnight at 150$^{\circ}$ C to remove the contaminants and to improve the diode quality as all fabrication steps were done subsequent to the Gr/Si interface formation unlike in the case for exfoliated graphene devices.    

\begin{figure*}
\includegraphics[scale=0.32]{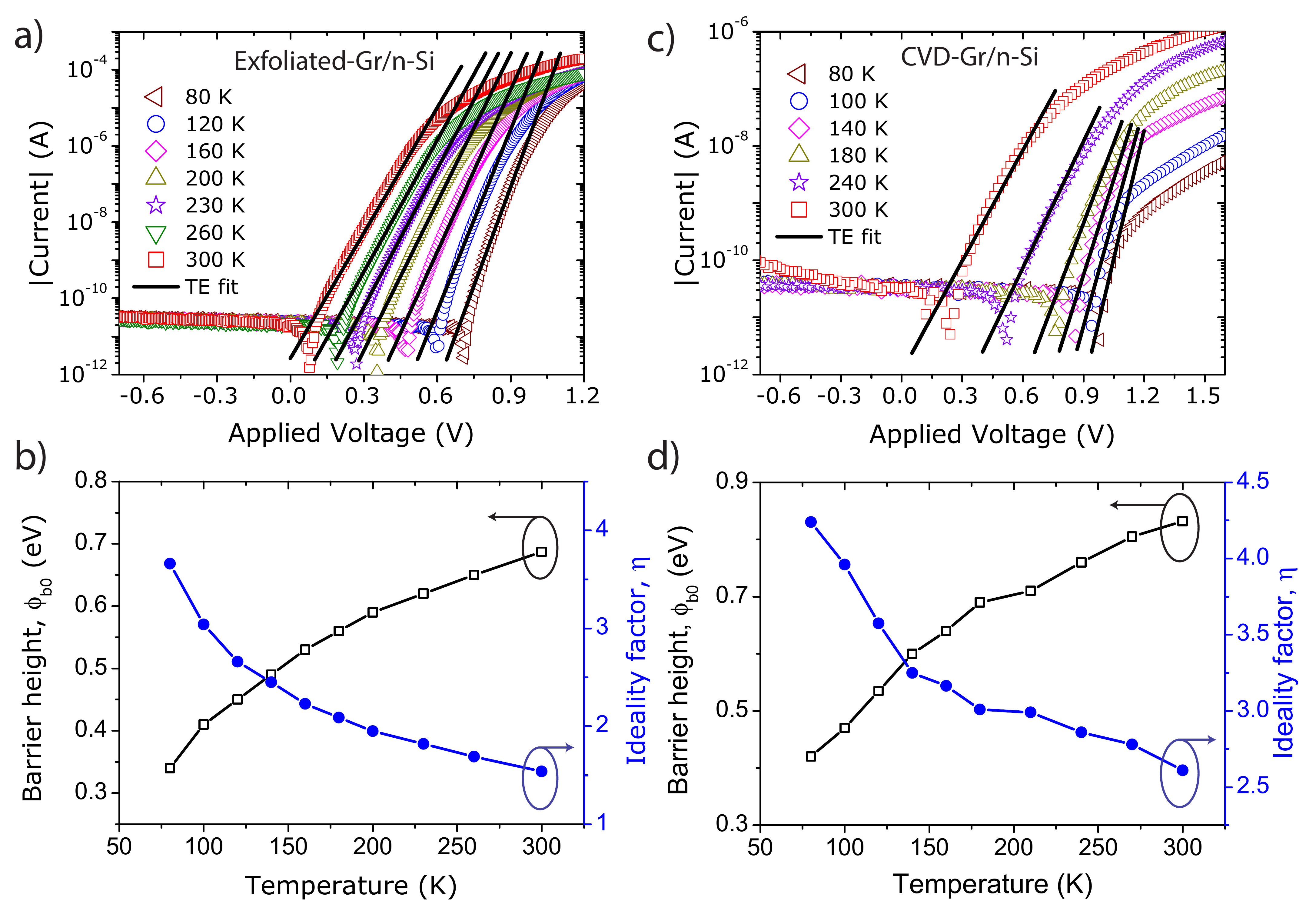}
\caption{\label{fig:Tdep_both} (a, c) Temperature dependent {\it I-V} measurements of the devices with exfoliated graphene (a) and CVD graphene (c) respectively. The solid lines are the linear fits using the thermionic emission equation. (b, d) Variation of the ideality factor ($\eta$) and zero-bias barrier height ($\phi_{b0}$) with temperatures for exfoliated (b) and CVD graphene (d) devices respectively.}
\end{figure*} 

The zero-bias Schottky barrier height (SBH), $\phi_{b0}$, and the ideality factor, $\eta$ at RT, are extracted for both diodes using thermionic emission (TE) theory according to the following equation
\begin{equation}
I=A^{**}AT^2\exp\left(-\frac{q\phi_{b0}}{k_BT}\right)\left[\exp\left(\frac{qV}{\eta k_BT}\right)-1\right],
\end{equation} 
where $q$ is the electronic charge, $k_B$ is the Boltzmann constant, $A$ is the Gr/Si interface area, $T$ is the temperature, $A^{**}$ is the effective Richardson constant which is 110 A cm$^{-2}$ K$^{-2}$ for an n-type Si(100) substrate. Extracted $\phi_{b0}$ and $\eta$ are 0.69 $\pm$ 0.01 eV and 1.46 $\pm$ 0.04 for the exfoliated-graphene [see Fig.~\ref{fig:RT_I-V}(b)] and 0.83 $\pm$ 0.02 eV and 2.53 $\pm$ 0.09 for the CVD-graphene [see Fig.~\ref{fig:RT_I-V}(c)] devices respectively. The ideality factor, $\eta$, is a quantity which determines the quality of the diode and can be written as,
$\eta=\frac{q}{k_BT}\frac{I}{\frac{\partial I}{\partial V}}$.
An ideal diode ($\eta$=1) represents purely TE process. In general, the Schottky barrier height represents the difference between the graphene work function ($W_{Gr}$) and the electron affinity of Si ($\chi$) i.e., $W_{Gr}$ - $\chi$ = (4.8 - 4.05) eV = 0.75 eV \cite{Solarcell}. We note however that the work function of graphene is reported to vary from 4.4 to 4.8 eV (Refs. 12 and 13) due to unintentional doping of graphene, thus leading to a wide distribution of SBH at different Gr/n-Si interfaces. Further, the presence of a thin layer of oxide at the interface during fabrication can result in an incomplete hydrogen passivation of the Si surface, creating interface states and dipoles with graphene. These result in an ideality factor greater than unity and a modification of the Schottky barrier \cite{Tung}, $\phi_{b0}$ = $W_{Gr}$ - $\chi$ + $qV_{\textit{int}}$, where $V_{\textit{int}}$ is the voltage drop due to interface dipoles. The difference in the barrier height and the ideality factor for the exfoliated-graphene and CVD-graphene devices are due to differences in the fabrication processes in addition to the differences in graphene quality which in principle can change the interface dipoles. To compare the quality of the graphene in our devices, we additionally fabricate graphene field effect transistors (FETs) on similar n-Si/SiO$_2$ substrates. We extract typical mobilities ($\mu$) at room temperature to be around 2200 cm$^{2}$V$^{-1}$s$^{-1}$ for exfoliated graphene and around 300 cm$^{2}$V$^{-1}$s$^{-1}$ for CVD-graphene respectively at similar charge carrier densities of 4$\times$10$^{12}$ cm$^{-2}$ for both and find them to be p-doped. We note that exfoliated graphene has better mobility ($\mu = \sigma/nq$) associated with high conductivity ($\sigma$) than CVD-graphene for similar carrier densities.    

In case of a non-ideal diode ($\eta$$>$1), there might be image force lowering of the barrier along with other transport channels such as thermionic field emission (TFE) and direct tunneling in addition to TE, which will modify the barrier height. As the doping concentration of n-Si is $10^{15}$ cm$^{-3}$, we can safely rule out direct tunneling,  as the calculated width of the depletion layer, $W_d$ ($>$1 $\mu$m) is too large at RT \cite{Chand}. For a better understanding of the transport characteristics, we also perform temperature dependence of {\it I-V} for both diodes. 

The {\it I-V} characteristics of the exfoliated Gr/n-Si Schottky diode in the temperature range 80 - 300 K is shown in Fig.~\ref{fig:Tdep_both}(a). The forward bias $\log(I)${\it-V} plots show a clear linear regime for the entire temperature range. A gradual increase of the onset voltage ($\approx$ forward-voltage at which the current starts to increase sharply) is observed from $\sim$0.1 V (300 K) to $\sim$0.8 V (80 K) with a decrease in temperature. This is because of the decrease in the thermal energy ($k_B$T) of electrons [from 26 meV (at 300 K) to 7 meV (at 80 K)]. The values of $\phi_{b0}$ and $\eta$ are extracted by fitting the linear part of the forward current (by Eq.~1) for all temperatures and plotted as a function of temperature in Fig.~\ref{fig:Tdep_both}(b). With decreasing temperature, the extracted ideality factor is found to increase (from 1.54 to 3.66), while the zero-bias barrier height decreases (from 0.69 to 0.34 eV).

Similar measurements are also obtained for CVD Gr/n-Si device and are plotted in Fig.~\ref{fig:Tdep_both}(c). Here too, we observe a gradual shift in the onset voltage towards higher bias when lowering the temperature which is consistent with the decrease in thermal energy of the electrons. However, the forward saturation current is now limited by the graphene quality and the quality of the Gr/Si interface as compared to the exfoliated graphene devices. A linear fit can be done for only few measurement points. The extracted $\phi_{b0}$ and $\eta$ are plotted in Fig.~\ref{fig:Tdep_both}(d) as a function of temperature. A similar trend as observed for exfoliated graphene device is found here with $\eta$ increasing from 2.61 to 4.24 and $\phi_{b0}$ decreasing from 0.83 to 0.42 eV upon decreasing temperature. The above results can be understood by considering the current transport across the Gr/n-Si interface to be a temperature activated process. At low temperatures, for inhomogeneous Gr/Si interfaces, transport is dominated by the low Schottky barrier patches while with increasing temperature, more electrons have sufficient energy to surmount the higher barriers and the extracted effective barrier increases. Although reports of such temperature dependence of $\phi_{b0}$ and $\eta$ exist in other metal/semiconductor (M/S) interfaces \cite{Gammon, Chand, Hubers}, no clear explanation other than the consideration of an inhomogeneous distribution of barrier heights has been proposed \cite{Tung2}. For Gr/Si interfaces, inhomogeneity in barrier heights arise due to local modification of graphene work function by the non-uniform interface charge distribution caused by unavoidable potential fluctuations due to ripples \cite{Rajput} and/or formation of graphene grain boundaries \cite{Tsen} etc.

\begin{figure}
\includegraphics[scale=0.29]{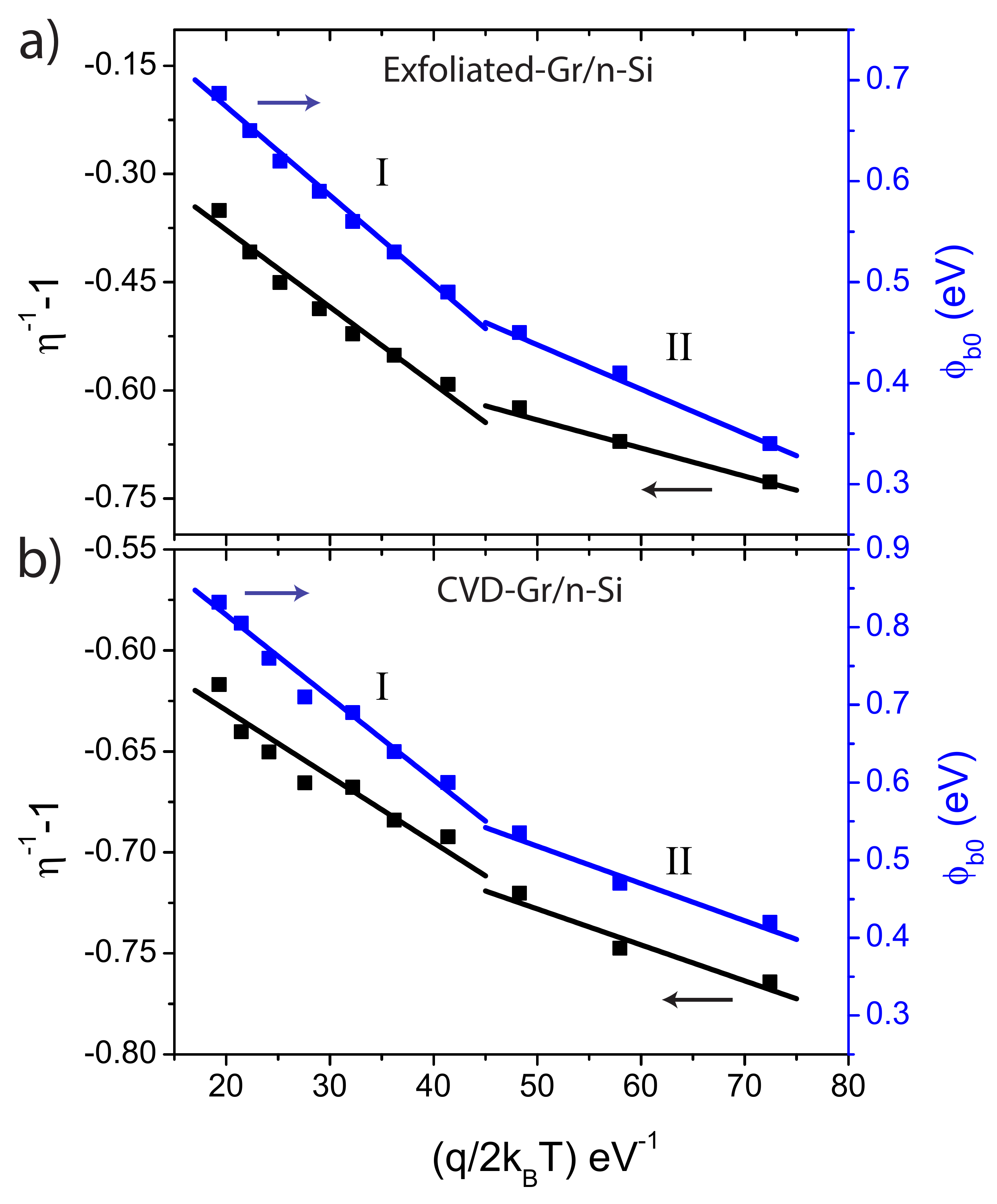}
\caption{\label{fig:T_phi_n} Zero-bias barrier height and the ideality factor versus $1/2k_BT$ curves for exfoliated-graphene/Si Schottky diode (a) and CVD-graphene/Si Schottky diode (b), respectively. Fits show two clearly observable temperature dependent regimes (I and II) for both diodes.} 
\end{figure}

A quantitative analysis of the mean barrier heights are further estimated by the Werner-G\"{u}ttler model \cite{Werner}, which takes into account the inhomogeneous distribution of the measured zero-bias barrier height, $\phi_{b0}$. According to the model, the Gaussian distribution of $\phi_{b0}$ yields the following equation
\begin{equation}
\phi_{b0} = \bar{\phi}_{bm} - {\frac{q{\sigma_s}^2}{2k_BT}},
\end{equation} 
where $\bar{\phi}_{bm}$ is the mean barrier height, and $\sigma_s$ is the standard deviation of the barrier distribution. A plot of $\phi_{b0}$ vs. $\frac{q}{2k_BT}$ therefore should yield a straight line with an intercept and a slope corresponding to $\bar{\phi}_{bm}$ and ${\sigma_s}^2$ respectively. The corresponding temperature dependence of the ideality factor, $\eta$ becomes \cite{Werner}
\begin{equation}
{\frac{1}{\eta}-1} = \rho_{2} - {\frac{q{\rho_{3}}}{2k_BT}},
\end{equation}
where $\rho_{2}$ (dimensionless) and $\rho_{3}$ are temperature independent voltage coefficients which quantify the voltage deformation of the barrier height distribution. Similarly, the plot of $(\frac{1}{\eta}-1)$ vs. $\frac{q}{2k_BT}$ should show a linear dependence with an intercept and a slope which depend on the voltage coefficients $\rho_{2}$ and $\rho_{3}$ respectively.

Figure 5 presents a combined plot of $(\frac{1}{\eta}-1)$ on the left y-axis and $\phi_{b0}$ on the right y-axis vs. $\frac{q}{2k_BT}$ as common the x-axis. As can be seen from figure 5, the linear dependence according to equations (2) and (3) has two distinct temperature regimes (I and II) for both graphene diodes. The values obtained for $\bar{\phi}_{bm}$ and $\sigma_s$ after fitting by equation (2) are summarized in table~\ref{table:T_phi_n}. 
 
The standard deviation, $\sigma_s$, is a measure of the barrier inhomogeneity; a large value of $\sigma_s$ corresponds to a more inhomogeneous distribution of the Schottky barrier heights. The values of $\sigma_s$ are significant compared to the $\bar{\phi}_{bm}$, which indicates strong inhomogeneities at both the interfaces and relatively larger inhomogeneities for the diode with CVD graphene than exfoliated graphene, as expected. Most importantly, $\bar{\phi}_{bm}$ and $\sigma_s$ are smaller in regime-II than regime-I, which indicate the stabilization of the potential fluctuation at low temperatures. Furthermore, we analyze the temperature dependence of the ideality factor using equation (3) and calculate the voltage coefficients $\rho_{2}$ and $\rho_{3}$, which are also summarized in table~\ref{table:T_phi_n}. 

The linear dependence of the experimental results with the potential fluctuation model in two different temperature regimes suggests a broad distribution of the Schottky barrier heights in both cases.

\begin{table}
\centering
\caption{Extracted fit parameters from Fig.~\ref{fig:T_phi_n} for both diodes of exfoliated (Ex.) and CVD graphene.}
\label{table:T_phi_n}
\begin{tabular}{lccccc}
\hline\hline
 &  & $\bar{\phi}_{bm}$ \raisebox{.5ex}{(eV)} & \raisebox{.5ex}{$\sigma_s$ (meV)} & \raisebox{.5ex}{$\rho_2$} & \raisebox{.5ex}{$\rho_3$ (meV)} \\ [0.45ex]
\hline 
\multirow{2}{*}{Ex.} & I & 0.85$\pm$ 0.04 & 93$\pm$ 6 & -0.16$\pm$ 0.04 & 10.7$\pm$ 0.6 \\ 
  & II & 0.66$\pm$ 0.03& 66$\pm$ 4 & -0.45$\pm$ 0.05 & 3.9$\pm$ 0.4 \\ [0.6ex]
 \multirow{2}{*}{CVD} & I & 1.03$\pm$ 0.06 & 103$\pm$ 7 & -0.56$\pm$ 0.05 & 3.3$\pm$ 0.3 \\ 
  & II & 0.76$\pm$ 0.04 & 69$\pm$ 5 & -0.64$\pm$ 0.06 & 1.8$\pm$ 0.2\\ 
  \hline\hline
\end{tabular}
\end{table}
           
\section{\label{sec:level1}Conclusions}
Realization of an optimal Schottky interface for graphene on Si is quite challenging as the barrier height strongly depends on the graphene quality and the fabrication processes employed. Such interfaces, unlike other M/S interfaces are found to be thermally and chemically stable in all environments \cite{Chen, TongayPRX} and are relatively easy to fabricate. Gr/Si interfaces are, thus, of increasing research interest for integration in future electronic devices. We successfully fabricate such interfaces at ambient conditions and find the electrical characteristics for both exfoliated and CVD graphene devices, to be highly rectifying, with minimal reverse leakage current and rectification of more than $10^6$ for the exfoliated device. Although the exfoliated graphene device shows a larger current density in the forward bias as compared to the CVD graphene device, a temperature dependent analysis of the electrical characteristics for both the devices show a decrease in barrier height and increase in ideality factor at low temperature. This indicates the presence of inhomogeneities at the Gr/n-Si interface due to ripples \cite{Rajput}, relatively high conduction at the edge and graphene grain boundaries \cite{Tsen} etc., which are difficult to control during the fabrication process. The induced electric field in monolayer graphene at a Gr/Si interface is virtually unscreened (the screening length in graphite is about $\approx$ 0.5 nm) \cite{Geim}, unlike the case for standard M/S interfaces, and opens further prospects in utilizing such interfaces in diverse electronic devices.

\section{\label{sec:level1}Acknowledgments}
We thank H. M. de Roosz and J. G. Holstein for technical support and K. G. Rana and I. J. Vera-Marun for helpful discussions. We acknowledge financial support from the Netherlands Organization for Scientific Research NWO-VIDI program, the Zernike Institute for Advanced Materials and NanoNed program coordinated by the Dutch Ministry of Economic Affairs.


\end{document}